\documentclass[pre,aps,superscriptaddress,fleqn,showpacs,twocolumn]{revtex4}
\usepackage{amsmath,amssymb,multirow,graphicx}
\begin{document}

\title{Converting genetic network oscillations into somite spatial pattern}

\author{K.~I.~Mazzitello}
\affiliation{CONICET} \affiliation{Facultad de Ingenier\'{\i}a,
             Universidad Nacional de Mar del Plata,
         Argentina}

\author{C.~M.~Arizmendi}
\affiliation{Facultad de Ingenier\'{\i}a,
             Universidad Nacional de Mar del Plata,
         Argentina}

\author{H.~G.~E.~Hentschel}
\affiliation{Department of Physics, Emory University,
             USA}

\date{\today}

\begin{abstract}
In most vertebrate species, the body axis is generated by
the formation of repeated transient structures called somites. This
spatial periodicity in
somitogenesis has been related to the temporally sustained
oscillations in certain mRNAs and  their associated gene products
in the cells forming the presomatic mesoderm. The mechanism underlying these  oscillations have been
identified  as due to the delays
involved in the synthesis of mRNA and translation into protein molecules [J. Lewis,
Current Biol. {\bf 13}, 1398 (2003)]. In addition, in the zebrafish embryo intercellular Notch signalling
 couples these oscillators  and a longitudinal
positional information signal in the form of an Fgf8 gradient exists that could be used to
transform these coupled temporal oscillations into the observed spatial periodicity of
somites. Here we consider a simple model based on this known biology and study its
consequences for somitogenesis. Comparison is made with the known properties of
somite formation in the zebrafish embryo . We also study the effects of localized Fgf8 perturbations
on somite patterning.\\
\end{abstract}
\pacs{05.40.-a, 05.60.-k, 73.63.-b}
\maketitle


\section{Introduction}
\label{intro}
\begin{Large}\begin{Large}
\end{Large}\end{Large}Somites are transient structures that form a periodic
growing pattern starting from the head (anterior) and extending to the tail (posterior) of a developing vertebrate embryo which
ultimately give rise to both the segmented vertebral column and to
the musculature. Formation of somites is a rhythmic process
characteristic of the species at given temperature. For instance,
in the chicken embryo one pair of somites is formed every 90 min
at 37 $^\circ C$ and in the zebrafish one pair is formed roughly
every 30 min at 28 $^\circ C$. The origin
 of these intracellular oscillations have been identified as due
 to time delays in the transcription and translation of
$her1$ and $her7$ genes  in the case of zebrafish and the $hes7$ gene in the case of
mouse. The total number of somites
produced is conserved within a given species typically somewhere between
50 to 70 pairs of somites form on each side of the anterior-posterior (AP) axis, but can vary dramatically
 between species, thus the zebrafish develops 30 pairs of somites.  As new somites bud from the anterior end of the presomatic mesoderm (PSM) that extends  back to the primitive streak
 and tail bud, new
 cells are added at the posterior end by cell division from the tail bud keeping the size of the PSM constant
 as new somites are generated .

Somitogenesis is one of the  best studied process of pattern
formation in the developing embryo, and a number of models have been
proposed to address the mechanisms underlying the generation of such
periodic patterns \cite{Baker:2006, Aulhela:2006, Pourquie:2003,
Aulhela:2004, Palmeirim:1997, Rida:2004}: the Clock and Wavefront model
proposed by Coke and Zeeman in 1976 is perhaps best known, but others
include Meinhardt's Reaction Diffusion
model; and Stern's Cell Cycle model. All of them suppose that
oscillations of genes and gene products occur in the cells of the PSM
 from which the somites derive (Fig. \ref{somites}). The Clock and Wavefront model
postulates the existence of a longitudinal positional information
gradient down the axis of vertebrate embryos. This gradient
interacts with the cellular oscillator stopping the oscillations and
producing  a rapid
change in locomotory and adhesive behavior of cells when they form
somites \cite{Dale:2000, Drubrulle:2001}.  The oscillation in the
PSM is the somite clock, and the moving interfaces at
the anterior end of the PSM where the positional information reaches a critical value  is called the wavefront. It is the interaction beween the clock phase and the positional
information that controls somitogenesis. Here we suppose the positional
information is supplied by the concentration gradient of the gene product
Fgf8 released at the tail bud of the embryo and creating a linear gradient
of morphogen along the AP axis \cite{Dale:2000,Pourquie:2003}. Biologically the signal created by the
Fgf8 concentration interacts with the clock phase (the intracellular
concentration of a gene protein) leading to cell differentiation and somite
formation. At low Fgf8 concentration, cell arrangement becomes more compact and
the epithelialization process underlying somite formation begins.

Meinhardt's Reaction Diffusion model also incorporates the idea of an
oscillatory mechanism in the PSM cells producing an alteration
between cell states as in the Clock and Wavefront model. One
of the main aims of this model is to account for somite structure
itself which at even early stages after formation shows heterogeneities
between the anterior and posterior portion of an individual somite. To
account for this structure the reaction diffusion model postulates that
cells can be in one
of two possible states: anterior $a$ and posterior $p$ and switch
from one state to another until they reach a stable state in the
presence of a morphogen gradient forming a
spatial pattern such as  $aaappaaapp ...$, where a repeated segment
$aaapp$ constitutes a somite.

Finally Stern's Cell Cycle model assumes
that the oscillations in the cells arranged along the AP axis are in synchrony with
each other, but cells in the anterior PSM are further advanced in phase than cells in the posterior PSM.
Segmentation is hypothesised to
occur when cells reach a certain time window in their cell cycle.
The most important point raised by this model is the question of how cells are
synchronized biologically. It appears that the answer to this question lies
in the $deltaC$ gene that codes for the Notch ligand at the heart of the intercellular
Notch signalling pathway which when disrupted destroys cell cycle synchrony\cite{Lewis:2003}.
All these models raise important questions, but recent experiments on how Fgf8 signalling controls somite
boundary position and regulates the segmentation clock
\cite{Drubrulle:2001} appears most in accord with the clock and wavefront model.

 Based in the Clock and
Wavefront model combined with key ingredients of the known biology,
we have therefore constructed a deterministic one dimensional model and compare
the resulting spatial patterning with
the somites for zebrafish embryo.
We use in our model the recent proposal for the mechanism
producing the intracellular oscillations as due to delays in the synthesis of mRNA
focussing specifically on the zebrafish genes $her1$ and $her7$ and their gene products the protein molecules
Her1 and Her7 \cite{Lewis:2003}.

The paper is
outlined as follows: Section \ref{model} is devoted to the
description of the proposed model and contains the
results derived from it. The transcription in the synthesis of mRNA
is essentially a noisy process and is considered for haploid
cells in the section \ref{haploid}. In section
\ref{irregular}, we study the effect of a local perturbation of the
Fgf8 wavefront on the spatial pattern of somites. Finally a conclusion
and discussion is presented in \ref{conclusions}.\\

\section{Deterministic one dimensional model}
\label{model}

\noindent The model consists initially of a linear array of $N_{start}$
cells that represents the starting size of the PSM with intracellular oscillators driven
by the $her1$ and $her7$ genes and their associated gene product proteins .
The oscillations between nearest-neighbor cells
are coupled and synchronized via Notch signalling\cite{Lewis:8,
Lewis:9}. The cell-cell communication produced by means of the $deltaC$
gene product Notch crossing the cell membrane wall.\\

\begin{figure}[!ht]
\begin{center}
\includegraphics[width=8 cm]{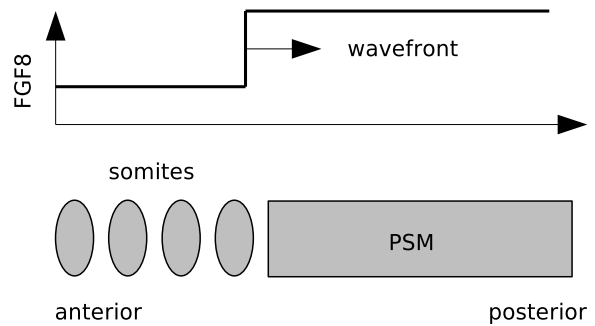}
\end{center}
\caption{An schematic illustration of somite formation within the Clock and
Wavefront model. Gene and gene product concentrations in the cells of PSM
experience temporal oscillations. In the top part of the diagram,
the positional information wavefront is illustrated together with the position of
the determination front which we assume represents the position at which the Fgf8 concentration
gradient falls below a critical value.  The somites form in an anterior to posterior order as the wavefront advances from head to tail.}
\label{somites}
\end{figure}

\noindent In each cell in the array the gene mRNA concentrations $m_k$  for the genes $k\equiv her1, her7, deltaC$ and their associated gene products the protein concentrations $p_k$ obey the sets of coupled kinetic equations (see \cite{Lewis:2003}):

\begin{eqnarray}
\frac{dp_{k}}{dt}(i,t)&=&a_{k}m_{k}(i,t-T_{p_{k}})-b_{k}p_{k}(i,t)\nonumber\\
\frac{dm_{k}}{dt}(i,t)&=&\frac{1}{nn}\sum_{i'=1}^{nn} f_k(p_{her1}(i,t-T_{m_k}), p_{her7}(i,t-T_{m_k}),\nonumber\\
&& p_{deltaC}(i',t-T_{m_k}))-c_{k}m_{k}(i,t)\nonumber\\
\label{eq:Lewis}
\end{eqnarray}
Here the symbol $i$ denotes the cell position in the linear array;  $i'$ goes
from 1 to $nn$ the number of near neighbor cells of $i$ ($nn$ can be 1
if $i$ is at an end or otherwise 2);  $T_{m_k}$ is the delay time from
initiation of transcription to export of the mature mRNA $m_k$  into the cytosol;
$T_{p_k}$ is the delay between the initiation of translation and the
emergence of the complete protein molecule $p_k$;   $a_k$ represents the protein $p_k$
synthesis rate per mRNA molecule;  $b_k$ is the rate of  protein $p_k$  degradation;
$c_k$ is  the rate of mRNA $m_k$ degradation;  while the function $f_k$ represents
the rate of production of new mRNA molecules $m_k$ that is given by the Michaelis-Menten
kinetics \cite{Lewis:2003}:

\begin{eqnarray}
&&f_k (p_{her1}(i,t'), p_{her7}(i,t'),p_{deltaC}(i',t'))=K_k \{ r0_k   \nonumber\\
&&+ rd_k\frac{\phi_{deltaC}(i',t')}{1+\phi_{deltaC}(i',t')}
+rh_k\frac{1}{1+\phi_{her1}(i,t')\phi_{her7}(i,t')}\nonumber\\
&&+rhd_k \frac{\phi_{deltaC}(i',t')}{1+\phi_{deltaC}(i',t')}\frac{1}{1+
\phi_{her1}(i,t')\phi_{her7}(i,t')}\}\nonumber\\
\label{eq:f_k}
\end{eqnarray}
where we used the notation $t'=t-T_{m_k}$; and $\phi_k(i,t)=p_k(i,t)/p^{crit}_k$. The
parameters $r0_k$, $rd_k$, $rh_k$, and $rhd_k$ that add up to
$1$, represent the relative weights of $k$ transcription that is
unregulated, regulated by $deltaC$ protein alone, regulated by $her$
protein alone, and regulated by $her$ and $deltaC$ proteins,
respectively. $p^{crit}_k$ is the critical number of molecules of Her1 or Her7
protein per cell, for inhibition of transcription if $k$ is $her1$  or
$her7$ respectively, while it represents the critical number of Notch molecules required for activation if $k$ is $deltaC$. The mRNA
concentrations for the kind different of genes decrease when the $k$
protein concentration are larger than their critical values $p^{crit}_k$ if $k$
is Her1 or Her7 or when the $deltaC$ protein Notch concentration is below
its critical value $p^{crit}_{deltaC}$. The concentrations of intercellular signalling
molecules in the cells are coupled between the nearest-neighbour $nn$ cells by means of
the function $f_k$ through the $deltaC$ protein Notch concentration of their neighboring
cells $p_{deltaC}(i',t-T_{m_k}))$, as it can be seen in the Eq. (\ref{eq:f_k}).\\

\begin{figure}[!ht]
\vspace{.3cm}
\begin{center}
\includegraphics[width=8 cm]{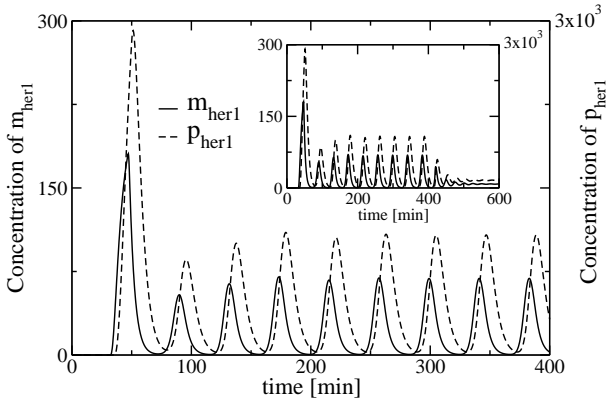}
\end{center}
\caption{Concentrations of  mRNA molecules of gene $her1$ and its associated gene product
 protein Her1 as a function of time for any cell of the linear array.
Sustained oscillations of both concentrations are obtained for the
following values of parameters involve in Eqs. (\ref{eq:Lewis}) and
(\ref{eq:f_k}):  $a_k=4.5$, $b_k=0.23$, $c_k=0.23$, $K_k=33,
r0_k=rd_k=0$, with $k=her1, her7, deltaC$, $rh_{her1, her7}=0$,
$rh_{deltaC}=1$, $rhd_{her1,her7}=1$, $rhd_{her1,her7}=1$,
$rhd_{deltaC}=0$, $p^{crit}_{her1,her7}=40$, $p^{crit}_{delta}=1000$,
$T_{m_{her1}}=12$, $T_{m_{her7}}=7.1$, $T_{m_{deltaC}}=16$,
$T_{p_{her1}}=2.8$, $T_{p_{her7}}=1.7$, $T_{p_{deltaC}}=20$ min.
(Inset) The oscillations are stopped when the corresponding cell is
reached by the somitogenesis wavefront. The concentrations of
protein and mRNA molecules of genes $her7$ and $deltaC$ have a similar
behavior as the concentrations of $m_{her1}$ and $p_{her1}$ shown in
this figure.} \label{temporal}
\end{figure}

\begin{figure}[!ht]
\begin{center}
\includegraphics[width=8 cm]{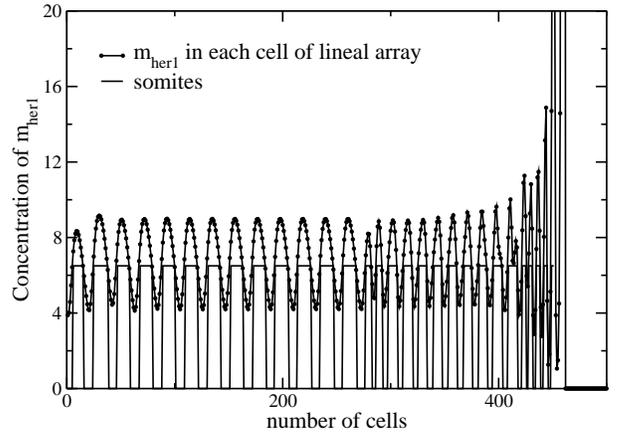}
\end{center}
\caption{Concentration of mRNA molecules of gene $her1$ as a
function of cell number in the somites. The
concentrations were calculated using Eqs. (\ref{eq:Lewis}) and
(\ref{eq:f_k}) and using the same parameter values as in the Fig.
\ref{temporal} (see caption). The concentration of $m_{her1}$
belonging to cells that were reached by the wavefront were frozen.
In our simulation the wavefront advances at a speed of one cell
every two minutes. We see the simulation at time $t=900$ minutes at the
end of somitogenesis in the zebrafish.
 At $t=900$ minutes 29 somites have formed. All the other concentrations of protein and mRNA
molecules in the cells (namely $p_{k}$ with k=Her1, Her7, Notch and
$m_{k}$ with $k=her7, deltaC$) have the same qualitative behaviour
as the concentration of $m_{her1}$.} \label{sin}
\end{figure}

The concentrations of both protein and mRNA molecules, $p_k$ and
$m_k$, are set equal to zero initially in all $N_{start}$ cells of the
linear array. The temporal evolution of  $p_k$ and $m_k$  are
calculated with Eqs. (\ref{eq:Lewis}). Sustained oscillations of
mRNA and proteins are obtained for parameters and time delays
$T_{m_k}$ and $T_{p_k}$ close to the zebrafish estimated ones as it
is shown in Fig. \ref{temporal} \cite{Lewis:2003}. The mRNA and
protein oscillations can occur without interactions between neighbor
cells, because the delays that take part in feedback loops are
intracellular. The interactions through $f_k$
connecting  neighbor cells are, however, crucial as they result in synchronous in phase oscillations in the linear
array representing the PSM as a whole to occur.\\

We suppose at time $t=0$ all the cells in the starting PSM to be oscillating  in phase,
and therefore we begin with the linear array of $N_{start}$ cells and integrate them for
the time necessary to obtain coherent synchronous oscillations. In this paper $N_{start}=280$, and in its synchronous
oscillating state it  represents the
time $t=0$ in our model. The growth
of the PSM   is simulated in our model by the adding
of new cells to the initial group of cells at one end of the linear array. This growth velocity $v_{growth}$ is a variable parameter in our model, and we chose to add cells at a rate of one cell every $5$ minutes in agreement with measured growth rates~\cite{Holley:2002}.  Thus, it generates
sequentially along the antero-posterior axis the necessary cells for
forming the future segments of the embryo: the somites that are created at
the anterior portion of the PSM, see Fig.~\ref{somites}.
In our simulations once the initial PSM has been obtained as a group
of cells oscillating in phase, the caudal motion of the wavefront
starts.   Again the wavefront velocity $v_{wave}$ is a parameter in the model that controls the rate at which cells reached by the wavefront stop oscillating and thus freezing the gene and gene product concentrations at their values when they meet the wavefront (see inset of Fig. \ref{temporal}).  The observed spatial shift in the phases of different cells is thus
obtained by the wavefront catching different phases
when passing through different cells, a biological implementation of the clock and wavefront model.

What value should we choose for $v_{wave}$? The simplest argument
would suggest $v_{wave}=v_{growth}$. This supposes that the FGF
creates a time indepedent distribution on a fast timescale compared
to growth, and therefore the wavefront is simply reading off a
critical value of the FGF concentration at which the intracellular
oscillator is frozen. If we chose this value, however, we would not
create around $30$ somites in $900$ minutes as we would
expect~\cite{Holley:2002}. Indeed,  it is important to note that the
regularity of the somite pattern and the size of somites is very
sensitive to changes in $N_{start}$  and $v_{wave}$. For example, if
$N_{start } =  20$ and $v_{wave}=v_{growth} \approx 5$ min per cell,
the number of somites is about 10 for 900 minutes. Here we report
instead on our simulations with $v_{wave} = 2$ minutes per cell.
Then when  the simulation finishes in the time $t=900$ minutes we
find  that $29$ somite pairs have been created. A wavefront
advancing at a rate of 1cell/2 minutes  creates a spatial wavelength
$\lambda = T_{cell} v_{wave} \approx 15$ cells in length. What is
the process of cell differentiation after the cell meets the
wavefront now depends on the biology. Here to be specific we have
assumed that when the $m_{her1}$ level is above its average value a
somite is formed. Averages of the concentrations of proteins and
mRNA of the frozen cells are calculated at this  final time. With
this choice a somite is roughly $6-7$ cells in length. On the other
hand the neighbor cells that have protein and mRNA concentrations
lower than average form the space between somites. With this choice
the remaining $8-9$ cells form an intersomite gap. Of course, it is
also possible to interpret the simulation as forming somites of
length $15$ cells with a differentiated structure between the
anterior and posterior portions of each somite. Now as the
simulation proceeds the PSM reduces in size. Indeed, for the
parameters chosen here $N_{start}=280$ cells, but $N_{end} =
10$cells, because while $180$ cells are added in the $900$ minutes
of the simulation at the posterior end of the PSM, a total of $450$
cells differentiate into new somites, It is not surprising therefore
that phase coherence is lost towards the end of somitogenesis, and
the spatial array of somites obtained, though approximately
periodic, decrease in size towards the tail (Fig. \ref{sin}). \\

\section{Stochastic Model}
\label{haploid}

Our simulations above assumed that no noise was present in the cells during transcription and translation, but as Lewis has
pointed out, the transcription step involved in protein synthesis is essentially a stochastic
process because of the small number of molecules involved in the cell \cite{Lewis:31}: A DNA molecule can randomly have a Her1,
Her7 or Delta Notch dimer bound to its regulatory site. When such a dimer does bind, transcription is forbidden; whereas if no protein is bound at a regulation site of the DNA molecule,
transcription occurs at the possible maximal rate.

The stochastic model for haploid cells is then constructed by taking the
 deterministic one dimensional model and considering the possible
states (bound and dissociated) of all regulatory sites of  the $her1$, $ her7$
and $deltaC$ genes as stochastic variables \cite{Lewis:2003}. A  DNA regulatory site
can have a Her1, Her7 or Delta Notch  dimer bound or not (i. e. either transcription forbidden or
maximally free of repression). A gene regulatory site then makes stochastic transitions between the bound and unbound states with a certain probability in any time interval.

 The protein and mRNA concentrations are given by
the Eqs. (\ref{eq:Lewis}) as in the  deterministic case, but the
function $f_k$ now describe a stochastic process and are therefore modified to:
\begin{eqnarray}
&&f_k (p_{her1}(i,t'), p_{her7}(i,t'),p_{delta}(i',t') )= K_k \{ r0_k  \nonumber\\
&&+rd_k [1-\xi_k^ {(D)}(i',t') ]+ rh_k \xi_k^{(H)}k(i,t')\nonumber\\
&&+rhd_k \eta _k^{(H)}(i,t') [1-\eta^{(D)}_k(i',t')] \} \nonumber\\
\label{eq:aploid}
\end{eqnarray}

\noindent where the parameters $r0_k$, $rd_k$, $rh_k$, and $rhd_k$
 were defined previously, below of Eq. (\ref{eq:f_k}). The functions
$\xi^{(D)}_k$, $\xi^{(H)}_k$, $\eta^{(H)}_k$ and $\eta^{(D)}_k$,
with $k=her1, her7, deltaC$, are random variables ($rv$) taking a
value $1$ when no protein is bound at the regulatory site  and $0$
when it is bound, with probabilities that depend on protein
concentrations in the cell $i$ and in its neighbor cell $i'$, at
time $t'=t-T_{m_k}$ when  synthesis of $m_k(t)$ begins. These
probabilities at a given time $t$ are thus conditioned by the values
the random variables take  at an earlier time $t'$. Specifically the
probability that any random variable ($rv$) takes value $1$ at the
time $t+\Delta t$ depends on the state of the appropriate regulatory
site at time $t$ (bound or unbound).

\begin{eqnarray}
p_{10}^{rv}(t+\Delta t) &=& 1-P^{rv}(t+\Delta t)|(P^{rv}(t)=0),{\mbox \;if\;regulatory}\nonumber\\
&&{\mbox site\;is\;bound\;at\;time\;t}\nonumber\\
p_{11}^{rv}(t+\Delta t) &=& P^{rv}(t+\Delta t)|(P^{rv}(t)=1),{\mbox \;if\;regulatory}\nonumber\\ 
&&{\mbox site\; is\; unbound\;at\; time\; t}\nonumber\\
\label{eq:p}
\end{eqnarray}

In Eq.~\ref{eq:p} $p_{\alpha \beta}^{rv}$ represents the probability that any random variable $rv$ has
value $\alpha$ at $t+\Delta t$ if it had value $\beta$ at $t$ ($rv$ can be
$\xi^{(D)}_k$, $\xi^{(H)}_k$, $\eta^{(H)}_k$ or $\eta^{(D)}_k$, and $\alpha =1 ,\beta = 0,1$); while
$P^{rv}(t)$ is the probability that the regulatory site is free of repression. This probability obeys the
master equation
 \begin{equation}\label{master}
dP^{rv}/dt = k_{off}^{rv}(1-P^{rv}) - k_{on}^{rv} P^{rv}.
\end{equation}
These kinetic coefficients are discussed in \cite{Lewis:2003} (Supplemental Data),
where $k_{on}^{rv}$ the rate constant for protein binding is a MIchaelis-Menten function of the
protein concentrations, while $k_{off}^{rv}$ is the rate constant for
dissociation of the repressor proteins from their  regulatory DNA binding sites (typically biologically  $k_{off}^{rv}\approx 1\;min^{-1}$ implying a mean lifetime of about
$1$ min for the repressor bound state).

 Solving the above differential
equation results in the forms $P^{rv}(t+\Delta t)=u^{rv}(1-exp[-v\Delta t]) + P^{rv}(t)exp[-v^{rv}\Delta t]$,
where $v^{rv}=(k_{off}^{rv}+k_{on}^{rv})$ and $u^{rv}=k_{off}^{rv}/v^{rv}$. Inserting these solutions in
Eq. (\ref{eq:p}) one obtains

\begin{eqnarray}
p_{10}^{rv}(t+\Delta t)&=&u^{rv} (1-e^{[-v^{rv}\Delta t]}),{\mbox \;if\;regulatory\;site\;is}\nonumber\\
&&{\mbox bound\; at\; time\; t}\nonumber\\
p_{11}^{rv}(t + \Delta t)&=&u^{rv}(1-e^{[-v^{rv}\Delta t]})+e^{[-v^{rv}\Delta t]},{\mbox \;if\;regulatory}\nonumber\\
&&{\mbox site\; is\; unbound\;at\; time\; t}\nonumber\\
\label{eq:p2}
\end{eqnarray}

\begin{figure}[!ht]
\begin{center}
\includegraphics[width=8 cm]{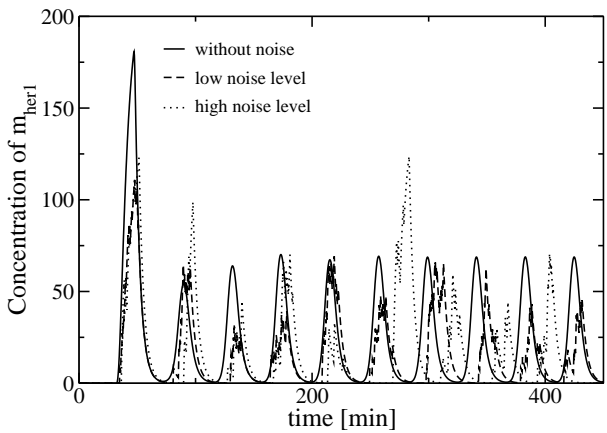}
\end{center}
\caption{Concentration of messager RNA molecules of gene $her1$ as a
function of time for any cell of the linear array. Comparison of
results obtained using the deterministic one dimensional model
described in section \ref{model} (solid line) and the stochastic
model for a low noise level (dashed line,
$k_{off}^{\eta^{(H)}_{her1/7}}=k_{off}^{\eta^{(D)}_{her1/7}}=10$ and
$k_{off}^{\xi^{(H)}_{deltaC}}=1$) and for a high noise level (dotted
line,
$k_{off}^{\eta^{(H)}_{her1/7}}=k_{off}^{\eta^{(D)}_{her1/7}}=1$ and
$k_{off}^{\xi^{(H)}_{deltaC}}=.1$). The parameter values  are the
same that in Fig. \ref{temporal}. The concentrations of protein and
message molecules have the same behavior as the concentration of
$m_{her1}$ shown in this figure.} \label{temporal2}
\end{figure}
In the same way, the probability that any random variable ($rv$)
takes value 0 at the time $t+\Delta t$ is given by

\begin{eqnarray}
p_{01}^{rv}(t+\Delta t)&=&(1-u^{rv})\left(1-e^{[-v^{rv}\Delta t]}\right), {\mbox\; if\; regulatory}\nonumber\\
&&{\mbox site\; is\; unbound\; at\; time\; t}\nonumber\\
p_{00}^{rv}(t+\Delta t)&=&1-u^{rv}\left(1-e^{[-v^{rv}\Delta t]}\right), {\mbox\; if\; regulatory}\nonumber\\
&&{\mbox site\; is\; bound\;at\; time\; t}\nonumber\\
\label{eq:p3}
\end{eqnarray}

\begin{figure}[!ht]
\vspace{.5cm}
\begin{center}
\includegraphics[width=8 cm]{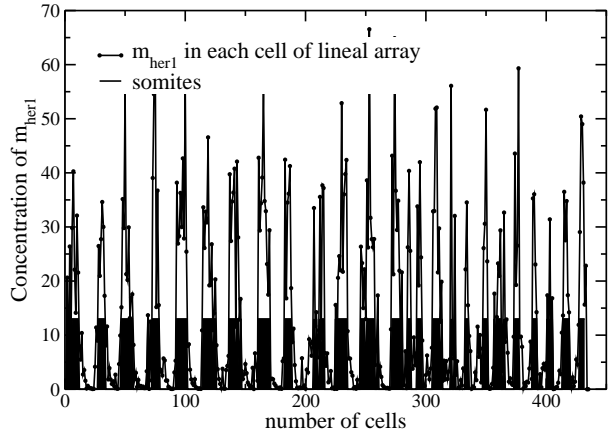}
\end{center}
\caption{Concentration of message molecules of gene $her1$ as a
function of the number of cell at the linear array. The
concentrations were calculated using the stochastic model for
haploid cells with a noise level given by
$k_{off}^{\eta^{(H)}_{her1/7}}=k_{off}^{\eta^{(D)}_{her1/7}}=5$ and
$k_{off}^{\xi^{(H)}_{deltaC}}=.5$, and assuming the rest of the
parameters equal to the used ones in the Fig. \ref{temporal}. The
wavefront advances at a speed of $2$ minutes by cell and $26$
somites are formed at $t=900$ min, in the simulation. All the other
concentrations of protein and message molecules in the cells ({\it
i.e.} $p_{k}$ with $k=her1, her7, deltaC$ and $m_{k}$ with $k=her7,
deltaC$) have similar behavior as the concentration of $m_{her1}$.}
\label{con2}
\end{figure}

Expressions (\ref{eq:p2}) and (\ref{eq:p3}) joined with the Eqs.
(\ref{eq:Lewis}) and (\ref{eq:aploid}) allow to calculate the
concentration of protein and mRNA molecules at anytime instant for
the linear array of cells considering transcription as a stochastic
process. There is a correspondence between this system with noise
and the deterministic system: statistical averages of the random
variables $\xi^{(D)}_k$, $\xi^{(H)}_k$, $\eta^{(H)}_k$ and
$\eta^{(D)}_k$ in the limit of large $k_{off}^{rv}$ and
$k_{on}^{rv}$ tend toward the behavior of the deterministic system.
In Fig. \ref{temporal2},  the concentration of $m_{her1}$ versus
time calculated using the deterministic model (detailed in the
section \ref{model}) and the model with noise for two different
levels of noise are compared. A random variability in the amplitude
and shape of individual oscillation peaks can be seen for both
results obtained with noise. In addition, as the amplitude of noise
increases the oscillations of $m_{her1}$ are not in phase with  the
deterministic model oscillations.\par Finally, a somite pattern for
the stochastic model is shown in Fig. \ref{con2}. The wavefront
advances at a speed of 2 minutes per cell freezing the oscillations
of protein and mRNA concentrations and the simulation finishes at
time t=900 minutes. Thus, a pattern of 26 somites is obtained with
the same constant values specified in the caption of Fig.
\ref{temporal} and also used in Fig. \ref{sin}. It is possible to
increase the number of somites decreasing the delays within the
range of allowed values (see \cite{Lewis:2003}).

\section{Perturbation of the somitogenesis wavefront}
\label{irregular}

The concentration of Fgf8 signalling  is high in the caudal PSM and
drops between the intermediate and the rostral PSM as indicated by
the top part of the Fig. \ref{somites}. In this paper we assume that
somite formation can begin when the Fgf8 signalling falls below a
critical level, and this level moves towards the caudal end of the
PSM with a constant velocity, the somitogenesis wavefront
$v_{wave}$. Note we assume this velocity is not the growth rate of
the PSM but in fact is faster $v_{wave}>v_{growth}$, we shall,
however, assume the same velocity in both the deterministic and
stochastic models. Now it is known that a transient manipulation of
the wavefront in zebrafish embryos alters the size of the somites
\cite{Holley:2002, Drubrulle:2001}: larger somites result when there
is transient inhibition of Fgf8 signalling, whereas smaller somites
result in the presence of transient activation.  Chemical inhibitor
and transplantation of Fgf8 beads are used for altering the level of
Fgf8 signalling that regulates the position of the wavefront within
the PSM \cite{Drubrulle:2001, Sawada:2001}.\par
\begin{figure}[!ht]
\vspace{.5cm}
\includegraphics[width=8 cm]{Fig6a.eps}
\end{figure}

\begin{figure}[!ht]
\vspace{1.cm}
\includegraphics[width=8 cm]{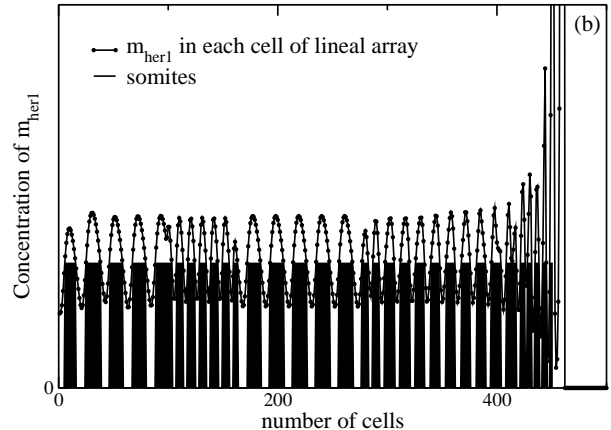}
\caption{Concentration of gene $her1$ mRNA as a
function of the number of cell in the linear array. The
concentrations were calculated using the deterministic version of the one
dimensional model and assuming the same parameter values as in the
Fig. \ref{temporal} (see caption) and \ref{sin}. The wavefront
advances at velocity of $2$ min per cell in both figures, but between
the cells $100$ and $160$ there is a perturbation in the Fgf8 levels resulting
in a change in the wavefront velocity: the velocity decreases to
$1$ min per cell in (a) and increase to $4$ min per cell in b). The
patterns of somites are altered by the perturbations (see Fig.
\ref{sin}).} \label{pert}
\end{figure}

Assuming that such perturbations in Fgf8 concentration directly
result in changes in the wavefront velocity, we will modify the
wavefront velocity in our model and analyze their consequences. The
perturbed pattern of the resulting somites depend on how the
wavefront velocity $v_{wave}$ changes. In Figure~\ref{pert} examples
of such perturbations are shown that have been obtained with the
deterministic one dimensional model by altering the wavefront
velocity (Fig. \ref{pert}). In each case the other parameters in the
model required for finding the protein and mRNA concentrations are
the same as in Fig. \ref{sin}. Specifically as in the unperturbed
system, the wavefront initially advances a rate of $2$ min. per cell
but between the cells $100$ and $160$ the wavefront is perturbed.
The velocity is increased to $1$ min per cell in the cell interval
$(100, 160)$ and larger somites are formed (see Fig. \ref{pert}(a))
in comparison with the results of Fig. \ref{sin}. When the velocity
is decreased to $4$ min per cell in the same interval of cells
smaller somites are formed (Fig. \ref{pert}(b)). In general
alterations of the wavefront velocity perturb the resulting pattern
somite in a grossly irregular manner.

\section{Conclusions}
\label{conclusions}  We have carried out simulations (see Figure \ref{con2}) using the deterministic version of our model that result in a growing approximately spatially periodic
sequence  of somites by combining the
clock and wavefront model with the temporal oscillation of gene transcription and translation
suggested by  Lewis~\cite{Lewis:2003} for the zebrafish embryo (Fig. \ref{sin}).
We also simulated the transient manipulation of the wavefront in zebrafish
embryos \cite{Holley:2002, Drubrulle:2001}. Larger somites are
formed when the wavefront velocity is locally increased
 whereas smaller somites results if there is a local decrease in the wavefront velocity (Fig. \ref{pert}).
Indeed, we believe that these are the first simulations that show in a biologically plausible manner the local variation of somite size due to  external perturbation of the Fgf8 concentration.

We also noted in
section \ref{haploid}, the consequences that gene regulation is in reality  a noisy
process which is likely to be crucial in the real developmental situation because of the small
number of intracellular molecules involved. Our simulations also raise a number of questions
for the biology os somitogenesis. First, as gene regulation is noisy it is important to study the
consequences of this noise for somite patterning. Second, the interplay between growth
at the caudal end of the PSM and the velocity of the somite development at the rostral end
is crucial for the number, size and shape of the resulting somites. It would be very useful there
to have more biological information of the relationship between these processes. Finally,
we  assume that the Fgf8 concentration
acts as a morphogen creating an intracellular signalling cascade that
ultimately controls transcription and translation of the gene products.
We do not have detailed information of this process and
simply assume that the Fgf8 acts as an on/off switch. It is certainly possible,
however, that a more complex dynamics may be involved and it would be useful
to have more information on Fgf8 signalling in this context.
In conclusion,  further developments in the model will
depend on knowledge of this biology and on  the pertinence to somite formation and differention on the phase of the oscillation cycle~\cite{Dale:2000}. \\


\end{document}